\documentclass[%
aip,
rsi,
amsmath,amssymb,
reprint,%
]{revtex4-1}

\usepackage{graphicx}
\usepackage{dcolumn}
\usepackage{bm}

\usepackage[utf8]{inputenc}
\usepackage[T1]{fontenc}
\usepackage{mathptmx}
\usepackage{etoolbox}

\usepackage{indentfirst}
\usepackage{newtxmath}

\usepackage{siunitx}
\NewCommandCopy\unitqty\qty
\DeclareSIUnit[quantity-product = \,]
\microkelvin{\text{$\mu$K}} 
\def\dsi{\vmathbb{i}}

\usepackage{physics}
\makeatletter
\def\@email#1#2{%
 \endgroup
 \patchcmd{\titleblock@produce}
  {\frontmatter@RRAPformat}
  {\frontmatter@RRAPformat{\produce@RRAP{*#1\href{mailto:#2}{#2}}}\frontmatter@RRAPformat}
  {}{}
}%
\makeatother
\begin{document}

\preprint{AIP/123-QED}

\title[ ]{A Radio-Frequency Emitter Design for the Low-Frequency Regime in Atomic Experiments}

\author{Yudong Wei}
\homepage{weiyd2017@pku.edu.cn}
\author{Zhongshu Hu}%
\author{Yajing Guo}
\author{Zhentian Qian}
\author{Shengjie Jin}
\homepage{jinshengjie@pku.edu.cn}
\affiliation{ 
  International Center for Quantum Materials, School of Physics, Peking University, Beijing 100871, China
}%
\affiliation{ 
  Hefei National Laboratory, Hefei 230088, China
}%
\author{Xuzong Chen}
\affiliation{%
School of Electronics Engineering and Computer Science, Peking University, Beijing 100871, China
}%
\author{Xiong-jun Liu}
\affiliation{ 
  International Center for Quantum Materials, School of Physics, Peking University, Beijing 100871, China
}%
\affiliation{ 
  Hefei National Laboratory, Hefei 230088, China
}%
\affiliation{
  International Quantum Academy, Shenzhen 518048, China
}



\begin{abstract}
Radio-frequency (RF) control is a key technique in cold atom experiments. We present a compact and efficient RF circuit based on a capacitive transformer network, where a low-frequency coil operating up to 30\,MHz serves as both an intrinsic inductor and a power-sharing element. 
The design enables high current delivery and flexible impedance matching across a wide frequency range. We integrate both broadband and narrowband RF networks into a unified configuration that overcomes the geometric constraints imposed by the metallic chamber.
In evaporative cooling, the broadband network allows a reduction of the applied RF input power from $14.7$\,dBW to $-3.5$\,dBW, owing to its non-zero coil current even at ultra-low frequencies. This feature enables the Bose-Fermi mixture to be cooled below 10\,$\mu$K. In a Landau-Zener protocol, the coil driven by the narrowband network transfers 80\% of rubidium atoms from $\ket{F=2, m_F=2}$ to $\ket{2, -2}$ in 1 millisecond, achieving a Rabi frequency of approximately 9\,kHz at an input power of 0.1\,dBW. 


\end{abstract}

\maketitle

\section{\label{sec:level1}
Introduction}

In experiments focused on preparing ultracold alkali metal atoms and molecules, a common approach involves trapping atomic ensembles at microkelvin ($\mu$K) in a magneto-optical trap (MOT), followed by confinement in a quadrupole magnetic trap. Simultaneously, an RF field is employed to remove high-energy atoms from the tail of the Maxwell-Boltzmann distribution by transferring them to high-field-seeking states, thereby forcing their ejection\cite{KETTERLE1996,Modugno2001,Bouyer2002}. Typically, when the timescale of radio frequency ramp\text{-}down is longer than the thermal equilibration time of the atomic ensemble, this process can lower the temperature of the atoms by an order of magnitude or more, providing a foundation for further evaporative cooling in an optical dipole trap. If the phase-space density is sufficiently high, the atomic ensemble can be cooled directly to reach quantum degeneracy\cite{Anderso1995,Davis1995,DeMarco1999}.


For alkali metals, two typical frequency regimes are used for evaporation in a quadrupole magnetic trap. The first involves RF transitions between sublevels within the same hyperfine manifold, while the second employs microwaves to transfer atoms to another hyperfine level.
The RF option, which generally involves scanning from 30\,MHz down to a few hundred kHz, is relatively easy to generate. However, the associated emitter network\footnote{We use ``network'' to describe the entire closed-loop system including source and load. By contrast, ``circuit'' refers to the unloaded passive components only.} often exhibits a high loaded quality factor $Q_L$, which narrows the impedance-matching bandwidth. As a result, the coil current becomes strongly frequency dependent, and the resulting variation in RF field strength leads to inefficient evaporative cooling as well as increased electromagnetic interference and unnecessary atom loss\cite{DingDing2024,Gehm2003,Roati2002a,Wang2011,stoferlet2005}.
In contrast, the microwave option enables shorter evaporation times\cite{Hachmann2022}, but it requires specialized equipment and imposes much stricter impedance-matching requirements on the antenna\footnote{We refer to any device that emits microwaves (on the order of gigahertz) as an ``antenna''.}, such as careful cabling and PCB layout design\cite{ELALAOUY2024,Firdaus2020}.
Furthermore, an additional microwave frequency must be applied during evaporation to remove certain intermediate states that would otherwise increase the probability of harmful inelastic collisions\cite{Modugno2002,DeMarco2019}.

In this paper, we present a simple yet effective design for the matching network for a low-frequency coil inspired by Ref.~\onlinecite{Scazza2021}, where a virtual load is used to improve matching and deliver higher current to the coil. However, this approach is limited by the large capacitive reactance and by the impractical length of the required quarter-wave transformer at ultra-low frequencies. In our version, the coil is treated as a built-in inductive element, forming part of a capacitive transformer network (CTN). The virtual load is embedded in the CTN as a passive termination to dissipate excess power, enabling a flexible and stable matching bandwidth. This design delivers several times more current to the coil than the conventional series-to-load scheme (i.e., coil in series with a resistive load) used in evaporative cooling\cite{Cohen2021}.

The design has several advantages. First, the coil branch is connected to the network in the simplest way, with only two soldering points, significantly reducing transmission losses. Second, using a virtual load to match the source resistance simplifies the matching process, promoting a low reflection coefficient while streamlining impedance adjustments. 
Additionally, as the matching network operates as a low-pass resonant circuit, the RF coil can maintain non-zero current even at ultra-low frequencies. These advantages make the design well suited for evaporative experiments requiring broadband matching, as well as internal state manipulations that demand high peak currents.

The CTN design enabled us to reduce the temperature of Rubidium-87 and Potassium-40 ($^{\text{87}}\text{Rb}-^{\text{40}}\text{K}$) mixtures trapped in a quadrupole magnetic trap from \unitqty{260(12)}{\microkelvin} to \unitqty{9(2)}{\microkelvin} within 10 seconds, while preserving the number of $^{\text{40}}\text{K}$ atoms with minimal loss. Subsequently, after transferring the mixtures into an optical dipole trap,  we achieved double quantum degeneracy, starting from a MOT containing only 2 million $^{\text{40}}\text{K}$ atoms. We also tested the narrowband version of the coil, which achieved a Rabi frequency of approximately \unitqty{9}{\kilo\hertz} between Zeeman sublevels of Rb atoms, with an input power of only 0.1\,dBW. The influence of the coil material and geometry on the performance of the design is also discussed in this work.

\section{Design and configuration}\label{sec2}
We first examine commonly used impedance transformer networks.
Among them, the L\text{-}type matching network (LTN) is widely adopted due to its simplicity, reliability, low loss, and cost-effectiveness\cite{Balanis2016} (Fig.~\ref{fig1}(b)). However, when applied to an emitter operating in the low-frequency regime, several issues arise. Let the complex impedance of the coil be $R_{\rm C} + \dsi X_{\rm C}$ and the source resistance be $R_{\rm S}$. In typical atomic experiments, the condition $R_{\rm S} \gg R_{\rm C}$ is often met, leading to a loaded quality factor for the network approximately given by $Q_L \sim X_{\rm C} /\qty(2 R_{\rm C})$ (see Appendix~\ref{appendixSecA}). We emphasize that the loaded quality factor $Q_L$ of the network, rather than the unloaded $Q_U$, is the relevant parameter governing the impedance-matching bandwidth, such as the width of the $\abs{S_{11}}$ dip. Since the reactance $X_{\rm C}$ is dominated by inductance even at the lowest frequencies, the resulting $Q_L$ remains well above unity (see Fig.~\ref{fig1}(c)). A high $Q_L$ restricts the impedance-matching bandwidth and increases sensitivity to frequency deviations, which may result in substantial mismatch and pose risks to the driving electronics.

Other issues at low frequencies include the large mismatch between the source and coil resistances, which often requires impedance transformations exceeding an order of magnitude. Additionally, low-pass matching networks rely on inductive elements, whose resistance is often comparable to that of the coil, resulting in unavoidable power losses and electromagnetic radiation.
These issues also arise in other transformer circuits, such as magnetic transformers based on mutual inductance.

In our scheme, the CTN design naturally facilitates the operation of the coil at ultra-low frequencies. It provides flexible impedance matching across a broad frequency range while minimizing the need for additional inductive components. The coil itself serves as the intrinsic inductor of the network, with a virtual load placed at the terminal to match the source resistance. The resulting voltage division simplifies impedance matching and enables the system to handle higher input power levels.


\subsection{General properties} \label{sec2_A}
\begin{figure}
  \includegraphics[scale=0.320]{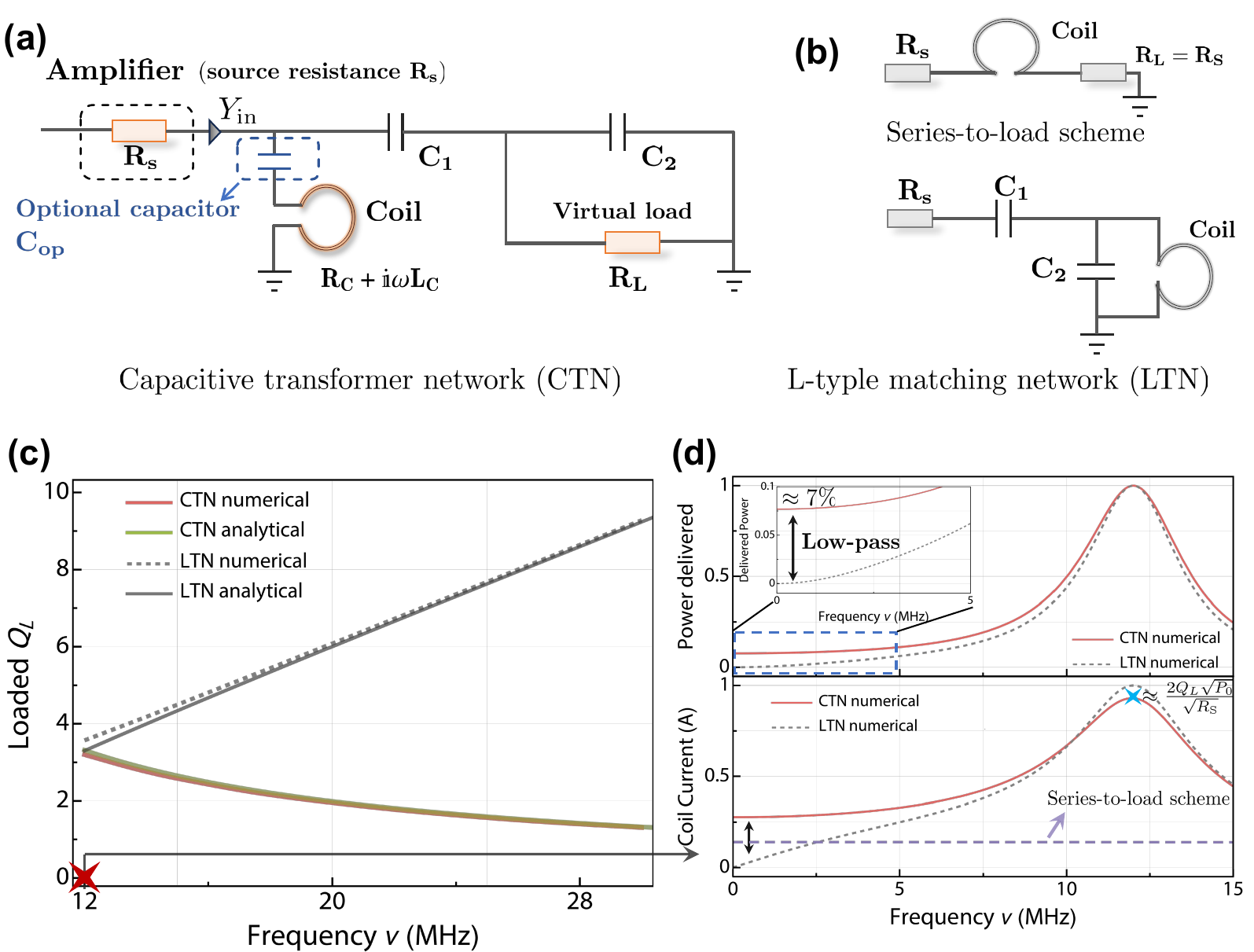}
  \caption{\label{fig1}Schematic of the CTN for impedance matching of a low-frequency RF coil, with $C_{\rm op}$ as an optional capacitor that is replaced by a wire unless otherwise specified. (a) The RF coil serves as the built-in inductor, canceling the reactance introduced by the capacitors. (b) Conventional designs for low-frequency RF networks, featuring a series-to-load scheme and a dual-capacitor LTN. (c) Fully numerically calculated $Q_L$ for the CTN (solid green line) is lower than that of the LTN (dashed line), indicating the CTN's potential for broadband impedance matching. The analytical result (green and gray line) agrees well with the numerical simulation. (d) Comparison of delivered power ratio from the amplifier, defined as $1 - |S_{11}|^2$, and coil current for the CTN and LTN at the resonant frequency of $12\,\mathrm{MHz}$ (indicated by the star in (c)). The LTN behaves as an open circuit at low frequencies due to its capacitive nature, while the CTN retains its inductive response, allowing continued power delivery in this regime. Parameters for (c)(d) are $L_{\rm C} = \unitqty{100}{\nano\henry}, R_{\rm C}=\unitqty{1}{\ohm}$, and $R_{\rm S} = R_{\rm L} =\unitqty{50}{\ohm}$, with $C_{1,2} = (0.67, 1.6)\,{\rm nF}$ for LTN and $C_{1,2}=(2.7, 4.7)\,{\rm nF}$ for CTN. The coil current in (d) is evaluated at a nominal input power of 0\,dBW.
  }
\end{figure}
The RF coil is modeled as a series RL element with impedance $Z_{\rm C}=R_{\rm C} + \dsi\omega L_{\rm C}$. Since we aim to optimize the current through the coil, the resistance must be taken into account. The basic structure of the CTN design is shown in Fig.~\ref{fig1}(a). The transformed input admittance can be written as $Y_{\rm in}=G_{\rm in}+\dsi B_{\rm in}$, where 
\begin{align} \label{eq1} 
  G_{\rm in}& \approx \frac{1}{\tilde{R}_{\rm C}}+\frac{1}{\tilde{R}_{\rm L}}, \nonumber\\ 
  B_{\rm in}&\approx \frac{\omega C_1C_2}{C_1+C_2}-\frac{1}{\omega L_{\rm C}}.
\end{align} 
We define $\tilde{R}_{\rm C} \equiv \qty[R_{\rm C}^2 + \qty(\omega L_{\rm C})^2] / R_{\rm C}$ and $\tilde{R}_{\rm L} \equiv R_{\rm L}\qty(C_1 + C_2)^2 / C_1^2$, with $\tilde{R}_{\rm L}$ denoting the approximate load resistance after an impedance transformation. Both $R_{\rm S}$ and $R_{\rm L}$ are typically \unitqty{50}{\ohm}. Matching at resonance is achieved when $G_{\rm in}=1/R_{\rm S}$ and $B_{\rm in}=0$. Eq.~(\ref{eq1}) provides a good approximation when $R_{\rm L}\gg 1/\qty(\omega C_{j})$ for $j=1,2$  and $R_{\rm C}\ll \omega L_{\rm C}$. In this limit the resonant frequency is
\begin{align} \label{eq2}
  \omega_0 \approx \frac{1}{\sqrt{L_{\rm C}C_{\rm eq}}},\qquad C_{\rm eq}\equiv\qty(\frac{1}{C_1}+\frac{1}{C_2})^{-1}.
\end{align}
For simplicity, we first consider the case without the optional capacitor $C_{\rm op}$.
Near resonance, when $\omega\approx\omega_0$ and $\omega + \omega_0 \gg \abs{\omega - \omega_0}$, the transfer function $A(\omega)$ of the whole network simplifies to\cite{Coleman2005}
\begin{align} \label{eq3}
  A(\omega) \approx \frac{R_{\rm in}}{R_{\rm in} + R_{\rm S}} \qty[\frac{1}{1 + \dsi \frac{2Q_L}{\omega_0} \qty(\omega - \omega_0)}]. 
\end{align}
Here, the loaded quality factor is $Q_L\equiv\frac{R_{\rm in} \parallel R_{\rm S}}{\omega_0 L_{\rm C}}$, and the input resistance
is $R_{\rm in} \equiv \tilde{R}_{\rm C} \parallel \tilde{R}_{\rm L}$, where ``$\parallel$'' denotes a parallel combination, i.e., $\qty(1/\tilde{R}_{\rm C}+1/\tilde{R}_{\rm L})^{-1}$.
Since $R_{\rm S}$ is generally fixed, the frequency-dependent $Q_L$ is directly related to the inductance of the coil. Small coils commonly used in experiments have inductances on the order of hundreds of \unit{\nano\henry}\cite{Barker2020,Herb2020,Scazza2021,Mispelter2015}. In the low-frequency regime, such a feature allows the $Q_L$ to remain within one order of magnitude, providing both physical flexibility and ease of realization.

We emphasize that the simple analytical forms of Eqs.~(\ref{eq1}) and (\ref{eq2}) are useful for quickly estimating component choices when the resonant frequency is below approximately 30\,MHz and $\tilde{R}_{\rm C} > R_{\rm S}$ (see Appendix\,\ref{appendixSecB}).
Most remarkably, the CTN without the optional capacitor $C_{\rm op}$ behaves as an intrinsic low-pass circuit from the coil's perspective. In the ultra-low-frequency limit, the capacitive branch becomes effectively an open circuit, while the coil branch, with an intrinsic resistance on the order of 1\,$\Omega$, still dissipates a non-negligible portion of the input power.  Specifically, even in this regime, at least $1-|S_{11}|^2 \approx 7.7\%$ of the input power reaches the coil, ensuring that a finite current continues to flow through it. As shown in Fig.~\ref{fig1}(d), we compare the CTN and the LTN, both tuned for resonance at 12\,MHz. This behavior ensures that RF evaporation remains effective in the final stage of cooling, regardless of how low the radio frequency becomes.


\subsection{Current characteristics}\label{sec:level12}
\begin{figure*}
\includegraphics[scale=0.54]{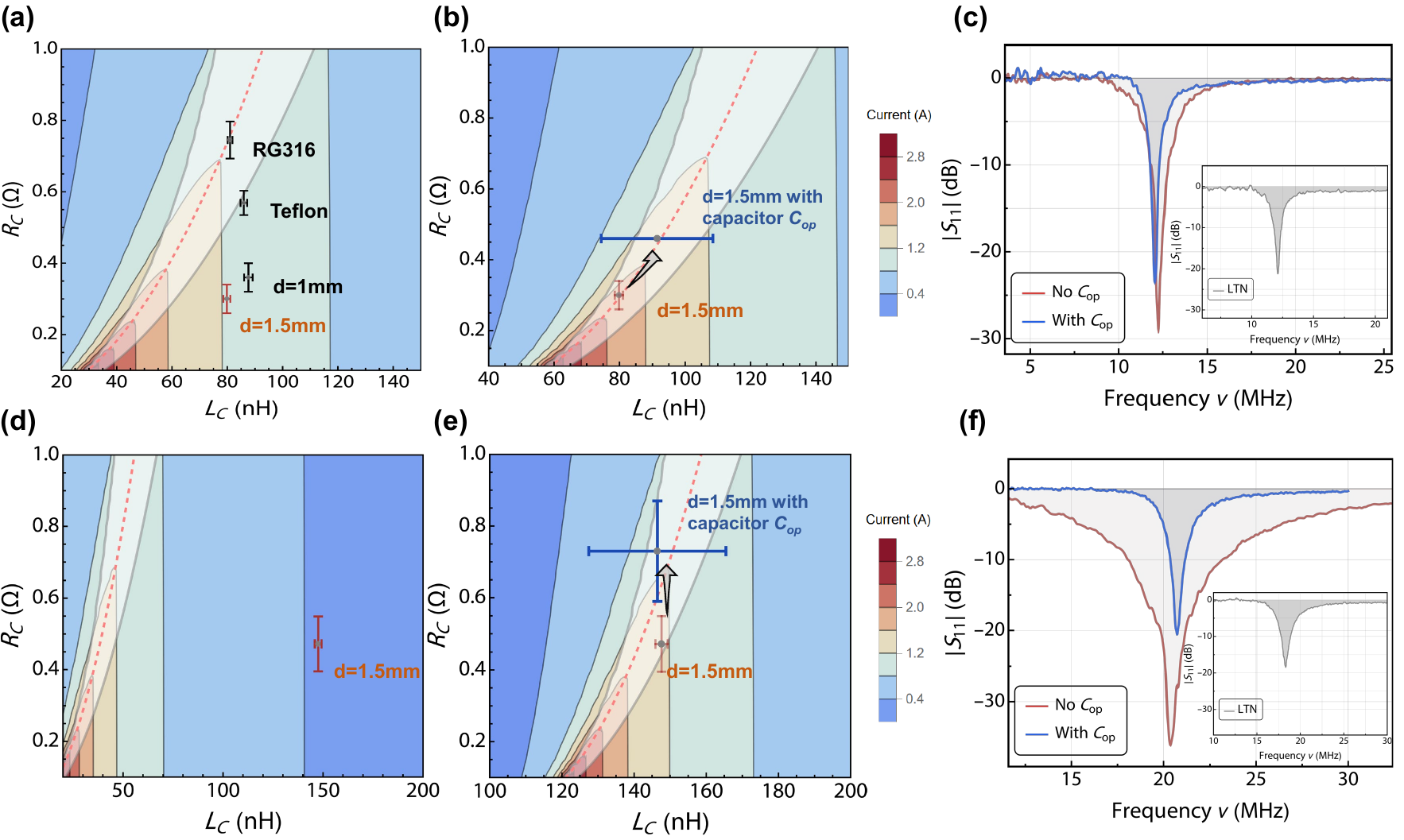}
\caption{\label{fig2} Effective current distributions for different inductances $L_{\rm C}$ and resistances $R_{\rm C}$ of the RF coil under resonant conditions at \unitqty{12}{\mega\hertz} (top row) and \unitqty{20.2}{\mega\hertz} (bottom row). The red dashed line denotes the boundary of feasible impedance matching (see text), where the coil receives the maximum power deliverable from the amplifier. The white region indicates the parameter range where at least 70\% of this maximum power is delivered to the coil. All impedance matching networks use the CTN configuration from Fig.~\ref{fig1}(a), unless otherwise specified.
(a) Comparison of four coil materials with different resistance and inductance values. 
(b) Adding an optional capacitor $C_{\rm op}$ shifts the current distribution, the resonance solution region, and the impedance matching region along the inductance axis. This adjustment enables the enameled copper wire to achieve higher current while maintaining good impedance matching.
(c)  $\abs{S_{11}}$ before and after the inclusion of $C_{\rm op}$, with comparison to our previous LTN configuration using the same enameled copper wire.
(d)-(f) Corresponding measurements and simulations at approximately \unitqty{20.2}{\mega\hertz}.
Capacitor values used in the CTN are:
For (a): $C_1 = \unitqty{5.6}{\nano\farad}$, $C_2 = \unitqty{3.3}{\nano\farad}$;
For (b): $C_1 = \unitqty{3.3}{\nano\farad}$, $C_2 = \unitqty{10}{\nano\farad}$, $C_{\rm op} = \unitqty{5.6}{\nano\farad}$;
For (d) and (e): $C_1 = \unitqty{15}{\nano\farad}$, $C_2 = \unitqty{0.36}{\nano\farad}$ and $C_1 = \unitqty{3.3}{\nano\farad}$, $C_2 = \unitqty{1.8}{\nano\farad}$, $C_{\rm op} = \unitqty{0.56}{\nano\farad}$, respectively.
All data points are obtained by fitting measured $\abs{S_{11}}$ curves and serve as approximate estimates of the parameters. $R_{\rm C}$ denotes the total resistance in the coil branch. The coils are circular single-turn loops with a diameter of \unitqty{4.1}{\centi\meter} in (a)-(c) and \unitqty{6}{\centi\meter} in the other panels.
The current distributions are calculated under an assumed 0\,dBW input power. 
}
\end{figure*}

One of the most interesting questions is whether it is possible to simultaneously achieve both a large peak current and a relatively broad impedance-matching bandwidth. Within the validity range of Eq.~(\ref{eq1}), we evaluate the delivered power $P_{\rm C}$ dissipated in the coil and $P_{\rm L}$ dissipated in the virtual load at exact resonance. Introducing $\beta=R_{\rm C}/R_{\rm L}$, we obtain:
\begin{align}
  \frac{P_{\rm C}}{P_{\rm L}} = \frac{\tilde{R}_{\rm L}}{\tilde{R}_{\rm C}} = \frac{4Q_L^2\beta}{4Q_L^2\qty(\beta^2-\beta)+1}.
\end{align}
In the practical regime $\beta\ll 1$, the effective peak current through the coil can be estimated as $I_{\rm peak} \approx 2Q_L \sqrt{P_0/R_{\rm S}}$. This corresponds to a current gain of approximately  $2Q_L$ compared to a series-to-load scheme. Here $P_0$ denotes the rated output power of the amplifier. This result also indicates that a lower $Q_L$ enhances the impedance-matching bandwidth, but at the cost of reduced peak current. The dependence of $I_{\rm peak}$ on loaded $Q_L$ thus provides a quantitative criterion for a trade-off between maximum current and impedance-matching bandwidth, which can be tuned by adjusting the inductance of the coil. Since the inductance depends not only on the coil geometry but also on its material and fabrication method, selecting different coil types offers flexibility in tuning the inductance to meet various experimental requirements.

\subsection{Maximizing current}\label{sec:level13}

We first examine how the current in a CTN can be optimized through the selection of different coil materials. Maximizing the current is essential for generating a strong RF magnetic field, which significantly enhances the atomic Rabi coupling, particularly in fixed-frequency schemes. Although the LTN can, in principle, deliver a larger current, as shown in Fig.~\ref{fig1}(d), this comes with several practical limitations, as previously discussed. 
In a high-power LTN, capacitors are especially vulnerable to failure, an issue we have repeatedly encountered in our laboratory. These considerations underscore the importance of optimizing the CTN  for maximal current delivery in practical settings.

According to the scheme described in Fig.~\ref{fig1}(a), we plot exact numerical results of the effective current through the coil for different values of lumped $R_{\rm C}$ and $L_{\rm C}$ at fixed frequencies of \unitqty{12}{\mega\hertz} and \unitqty{20.2}{\mega\hertz} under resonant conditions in Fig.~\ref{fig2}(a)(b) and (c)(d), respectively. The red dashed line represents the critical condition for $R_{\rm C,crit}$ and $L_{\rm C,crit}$, corresponding to $R_{\rm in} \approx \tilde{R}_{\rm C} = R_{\rm S}$. This implies that $\tilde{R}_{\rm L} \gg \tilde{R}_{\rm C}$, so that most of the power is delivered to the coil, resulting in the maximum achievable current.  
Moreover, to the left of the red dashed line, since $L_{\rm C}<L_{\rm C,crit}$, which leads to $\tilde{R}_{\rm C}<R_{\rm S}$, no solutions satisfy exact resonance (see Appendix\,\ref{appendixSecB}). Therefore we use the capacitors $C_{1,2}$ solved on the red dashed line for the plot\footnote{
Here, Fig.~\ref{fig2} is generated by fixing $R_{\rm C}$ and scanning $L_{\rm C}$ from large to small. An alternative approach would be to fix $L_{\rm C}$ and scan $ R_{\rm C}$ from small to large. The latter method may yield slightly higher currents to the left of the red dashed line. Since most experimental coil parameters lie on the right side of the red dashed line, both approaches give essentially identical current values in the relevant region. Near the red dashed line $\tilde{R}_{\rm C} = R_{\rm S}$, where $ \tilde{R}_{\rm L} \gg \tilde{R}_{\rm C}$, numerical solutions for $C_{1,2}$ may exhibit rapid variations. Nonetheless, this has little effect on the current distribution shown in Fig.~\ref{fig2}. }. 
The white area highlights the most efficient parameter range for the RF coil, where it captures over 70\% of the maximum output power from the amplifier.

For a narrowband emitter, it is essential that the parameters fall within the white region and that the coil have sufficiently low resistance. Without using an optional capacitor, the parameters can be modified by different materials, i.e., they effectively shift the values of $R_{\rm C}$ and $L_{\rm C}$. We consider common materials used in atomic experiments, including enameled copper wire with diameters of \unitqty{1}{\milli\meter} and  \unitqty{1.5}{\milli\meter}, Teflon-coated tinned wire, and the braided layer of RG316 cables.
We first use an LCR meter to obtain preliminary measurements of the inductance $L_{\rm C}$ and resistance $R_{\rm C}$ of the coils mounted on the holder but not yet soldered to the PCB. Based on these values, the required capacitances $C_{1,2}$ are numerically calculated using Eq.~(\ref{eq1}). After soldering four different coil materials into the CTN, we fit the measured $\abs{S_{11}}$ trace to extract their corresponding $R_{\rm C}$ and $L_{\rm C}$. The results are shown in Fig.~\ref{fig2}(a). While the inductance of these materials varies slightly for the same shape, their resistances differ significantly. The thicker enameled copper wire exhibits the smallest resistance. However, at low frequencies it falls outside the white region (to the right), thereby moving away from optimal impedance matching. RG316, on the other hand, has a higher resistance and lies exactly on the red dashed line, allowing for optimal impedance matching, while its higher resistance limits the increase in current. 

To further increase the peak current near the resonance, we place an optional capacitor in series with the \unitqty{1.5}{\milli\meter} enameled copper wire. This capacitor effectively shifts the current distribution pattern in Fig.~\ref{fig2}(a) along the inductive axis, resulting in the current distribution shown in Fig.~\ref{fig2}(b). The amount of this shift is determined by $1/\qty(\omega^2 C_{\rm op})$, which provides a convenient way of estimating the required capacitor value. Since PCB soldering typically introduces parasitic capacitance, we empirically increase the simulated $C_{\rm op}$ values by $5\text{--}10\%$ to better approximate the effective capacitance in the actual network.
In this case, the choice of $C_{1,2}$ is still guided by the formulas in Sec.~\ref{sec2_A}, provided the equivalent capacitance is redefined to include $ C_{\rm op} $ as $ C_{\rm eq} = \qty(1/C_1+1/C_2+1/C_{\rm op} )^{-1}$.
The measured $\abs{S_{11}}$ before and after adding $C_{\rm op}$ are shown in Fig.~\ref{fig2}(c). The addition of $C_{\rm op}$ results in a 25\% increase in coil current, at the cost of a narrower resonance dip. Conversely, placing $C_{\rm op}$ in series with the coil suppresses the low-frequency current response, making it more suitable for fixed-frequency control. 
A Lorentzian fit to the experimental data shows that the width of the good matching regime (where $\abs{S_{11}}<-10\,{\rm dB}$) decreases from \unitqty{1.06}{\mega\hertz} to \unitqty{0.57}{\mega\hertz}. Despite this reduction, the narrowband CTN design maintains a comparable performance to the LTN while also offering improved tolerance to higher input power.

Similarly, we tested a resonant frequency at \unitqty{20.2}{\mega\hertz}, confirming that the addition of the optional capacitor effectively maintains the coil current while improving impedance matching. Particularly, before adding $C_{\rm op}$, the design exhibits both a wide good matching regime of up to 5.4\,MHz and a deep minimum in $\abs{S_{11}}$ (Fig.~\ref{fig2}(f)), indicating strong and broadband impedance matching. Combined with the low-pass characteristics of the CTN in the ultra-low-frequency regime, these features make the design well suited for evaporative cooling.

\section{The Coils in Experiments}
We then validate the performance of the network designs in cold atom experiments, including evaporative cooling and Landau-Zener tunneling within $^{\text{87}}\text{Rb}-^{\text{40}}\text{K}$ mixtures confined in a magnetic trap. For evaporative cooling, Eq.~(\ref{eq2}) suggests that a relatively large inductance is required to reduce $Q_L$, thereby broadening the impedance-matching bandwidth. This ensures that the coil can maintain a fairly flat current response over a wide frequency range up to 30\,MHz. Naturally, this broader bandwidth results in a reduced peak current $I_{\rm peak}$. Considering the constraints of the coil geometry, we selected the braided shield of RG316 cable as the conductor. After installation, the coil exhibits $R_{\rm C} \approx 0.7\,\Omega$ and $L_{\rm C} \approx 175\,{\rm nH}$,  closely matching the red trace in Fig.~\ref{fig2}(f), which shows broadband $\abs{S_{11}}$ performance over the target frequency range.

For manipulating Zeeman sublevels within the same hyperfine manifold under weak magnetic fields, we target a frequency of 4\,MHz. 
At such low frequencies, exact resonance can still be achieved by increasing the coil inductance (see Appendix~\ref{appendixSecB}).
To maximize the current at resonance, it is necessary to minimize the coil resistance. We therefore selected enameled copper wire with a diameter of \unitqty{1.0}{\milli\meter}, resulting in $R_{\rm C} \approx 0.35\,\Omega$ and $L_{\rm C} \approx 275\,{\rm nH}$.

\begin{figure}
  \includegraphics[scale=0.320]{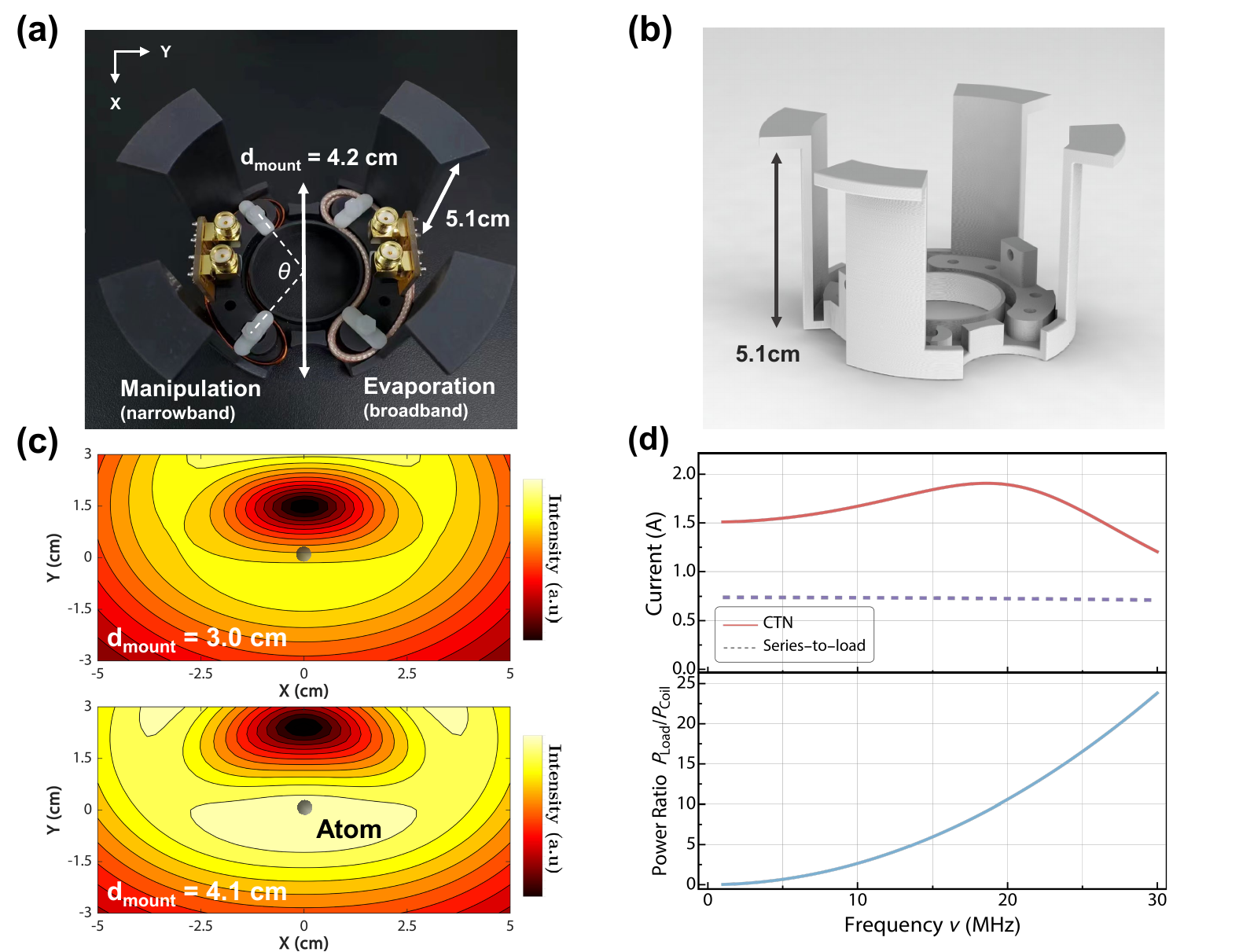}
  \caption{\label{fig3}Configuration of the two coils (made from RG316 and $d$=\unitqty{1.0}{\milli\meter} enameled copper wire) used for evaporative cooling and sublevel manipulation, with impedance matching measurements and simulations corresponding to the former (right) coil. 
  (a) Photograph of the two half-annular coils mounted on a 3D-printed resin holder. The holder is installed just above the upper viewport of the vacuum chamber. $d_{\rm mount}$ denotes the inner diameter of the holder and the outer diameter of the coil.
  (b) Engineering schematic of the coil holder.
  (c) Simulated horizontal RF magnetic field at the atomic position for an ideal conductor in vacuum. The dot indicates the relative location of the atomic cloud.
  (d) Simulated current and power through the evaporation coil as a function of frequency, based on the experimentally applied input power of 14.7\,dBW, with transmission losses taken into account. Impedance mismatch at ultra-low frequencies is not a concern, as the input RF power has already decreased by two orders of magnitude at these frequencies (see Fig.~\ref{fig4}(b)). The electronic components used are $R_{\rm C} \approx \unitqty{0.7}{\ohm}$, $L_{\rm C} \approx \unitqty{175}{\nano\henry}$, and dual capacitors $C_1 = \unitqty{4.7}{\nano\farad}$, $C_2 = \unitqty{0.3}{\nano\farad}$ for evaporation, and $R_{\rm C} \approx \unitqty{0.35}{\ohm}$, $L_{\rm C} \approx \unitqty{275}{\nano\henry}$, $C_1 = \unitqty{10}{\nano\farad}$, $C_2 = \unitqty{15}{\nano\farad}$ for sublevel manipulation.}
\end{figure}

A common choice for lab-made coil geometry is the circular shape\cite{Ciampini2006}. Instead, we chose a half-annular design\cite{Scazza2021} because it allows both the broadband and narrowband coils to be mounted symmetrically on the holder and positioned as close as possible to the atoms. This configuration also simplifies installation, as shown in Fig.~\ref{fig3}(a)(b).
The holder is mounted outside the upper viewport of the metallic chamber, with its bottom surface located approximately $d_0 = 4.1\,\mathrm{cm}$ above the atoms. 
Our simulations also show that, for an opening angle fixed at \( \theta \approx 110^\circ \), the coil generates the strongest horizontal magnetic field at the atomic position when its mounting diameter \( d_{\rm mount} \) is comparable to this distance, i.e., $ d_{\rm mount} \approx d_0$ (Fig.~\ref{fig3}(c)).

\subsection{The evaporation}

\begin{figure}
  \includegraphics[scale=0.320]{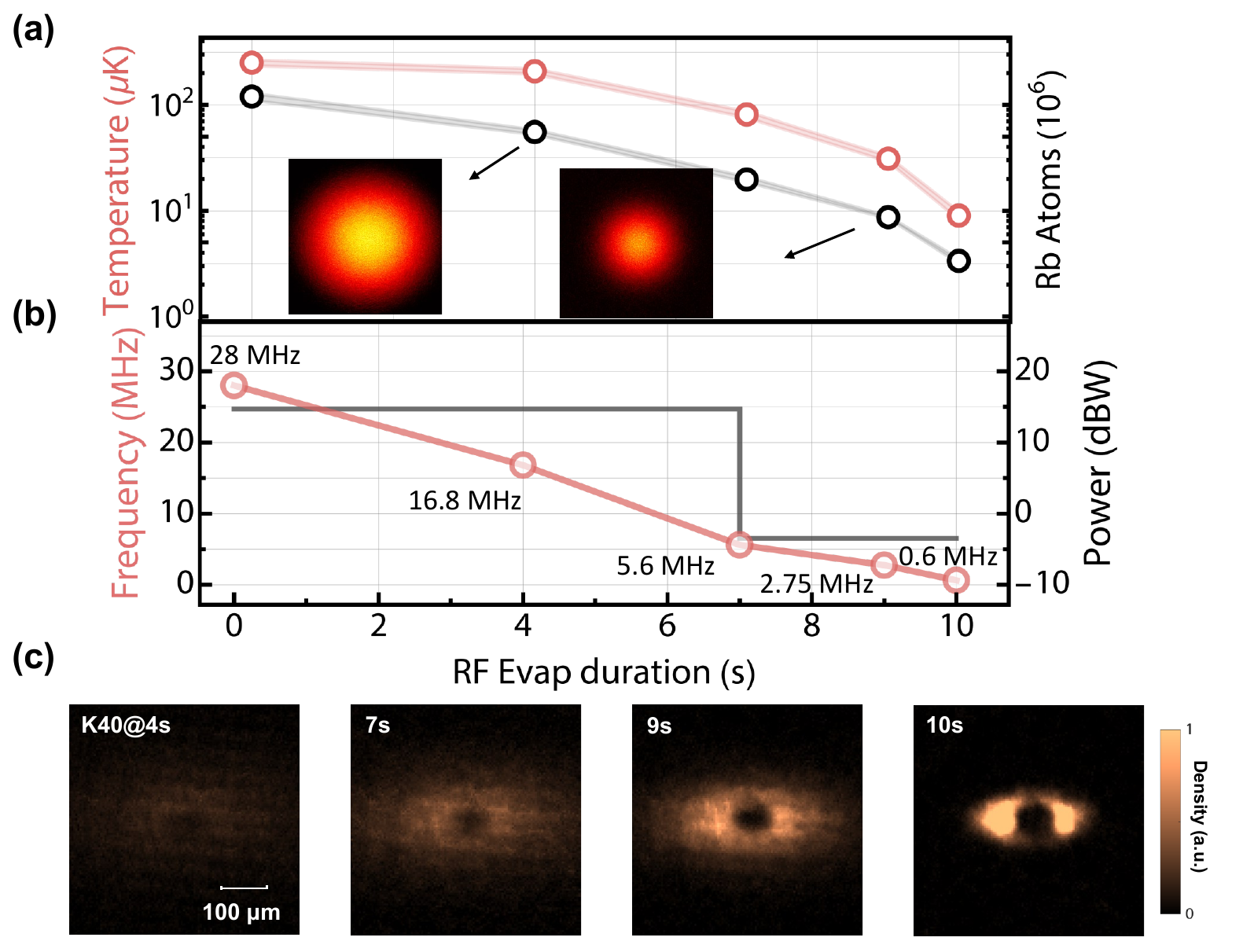}
  \caption{\label{fig4}RF evaporative cooling of a Bose-Fermi mixture in a quadrupole magnetic trap.  
  (a) The variation in $^\text{87}\text{Rb}$ atom number and temperature over four stages. The atom number decreases from $1.2 \times 10^7$ to $3.8 \times 10^5$, while the temperature drops from \unitqty{260(12)}{\microkelvin} to \unitqty{9(2)}{\microkelvin}. Shaded bands represent the standard deviation obtained from fitting. Individual error bars are smaller than the marker size. 
  (b) Radio frequency is decreased almost linearly from \unitqty{28}{\mega\hertz} to \unitqty{0.6}{\mega\hertz} over 10 seconds. The power was reduced by more than 18\,dB (from 14.7\,dBW to -3.5\,dBW) during the final two stages to minimize atom loss, since the coil maintains substantial current at ultra-low frequencies.
  (c) In situ absorption images of $^\text{40}\text{K}$ atoms at each stage reveal a significant increase in density. The plug beam has a waist of about \unitqty{40}{\mu\meter}, a power of \unitqty{800}{\milli\watt}, and a wavelength of \unitqty{760}{\nano\meter}.  The magnetic field gradient is held constant at 150\,G/cm throughout the evaporation sequence. }
\end{figure}

In the experiment, we loaded about $1.2\times10^7\,^{\text{87}}\text{Rb}$ atoms in the $\ket{F=2,\,m_F=2}$ state and about $2\times10^5\,^\text{40}\text{K}$ atoms in the $\ket{9/2,\,9/2}$ state into a 150$\,\text{G/cm}$ optical-plugged magnetic trap. The initial temperature of the $^\text{87}\text{Rb}$ atoms was around \unitqty{260}{\microkelvin}. We performed RF evaporation in four stages, gradually lowering the frequency and power while keeping the trap depth constant. After these stages, the temperature of the $^\text{87}\text{Rb}$ atoms markedly dropped to below \unitqty{10}{\microkelvin}. Meanwhile, the density of the $^\text{40}\text{K}$ atoms increased, with no noticeable loss in total atom count. 

Fig.~\ref{fig4}(a) shows the variation in atom number of Rb across different stages of the evaporation sequence, while Fig.~\ref{fig4}(b) presents the corresponding evolution of the radio frequency and input power as a function of evaporation time. The insets in (a) show the momentum distribution of Rb atoms after 10\,ms time-of-flight (TOF), taken at the first and third stages of evaporation sequence, with identical field of view for visual comparison. Meanwhile, the density of $^\text{40}\text{K}$ atoms gradually increased at each stage, as shown in Fig.~\ref{fig4}(c). Due to the small atom number and the segmented distribution of potassium caused by the plug beam, we were unable to accurately measure its temperature in the magnetic trap. Nevertheless, in the subsequent optical dipole trap, we confirmed that the temperatures of both species were consistent with each other\footnote{The transfer efficiency of fermionic potassium from the magnetic trap to the optical dipole trap was about 50\%, with losses primarily due to segmentation effects by the plug beam}. 
The gradual variation of coil current facilitates the RF evaporation process, particularly by simplifying the tuning of RF power and sweep rate compared to the LTN design. In the final two stages of evaporation, it becomes essential to reduce the RF input power to minimize RF-induced heating of the atoms.

The result also confirms the low-pass characteristics of our design. In contrast, typical LTN designs with dual capacitors often struggle to maintain an effective RF field during the final stages, even with a relatively large input power. The current evaporation sequence is likely limited by the magnetic field gradient of the quadrupole trap, which is only 50\text{--}70\% of that typically used in other experiments\cite{stoferlet2005,Bloom2014,Hachmann2022}. A higher gradient, or further optimization of the RF power trajectory, may lead to more efficient and faster evaporation.

\subsection{The Landau-Zener manipulation}
\begin{figure}
  \includegraphics[scale=0.320]{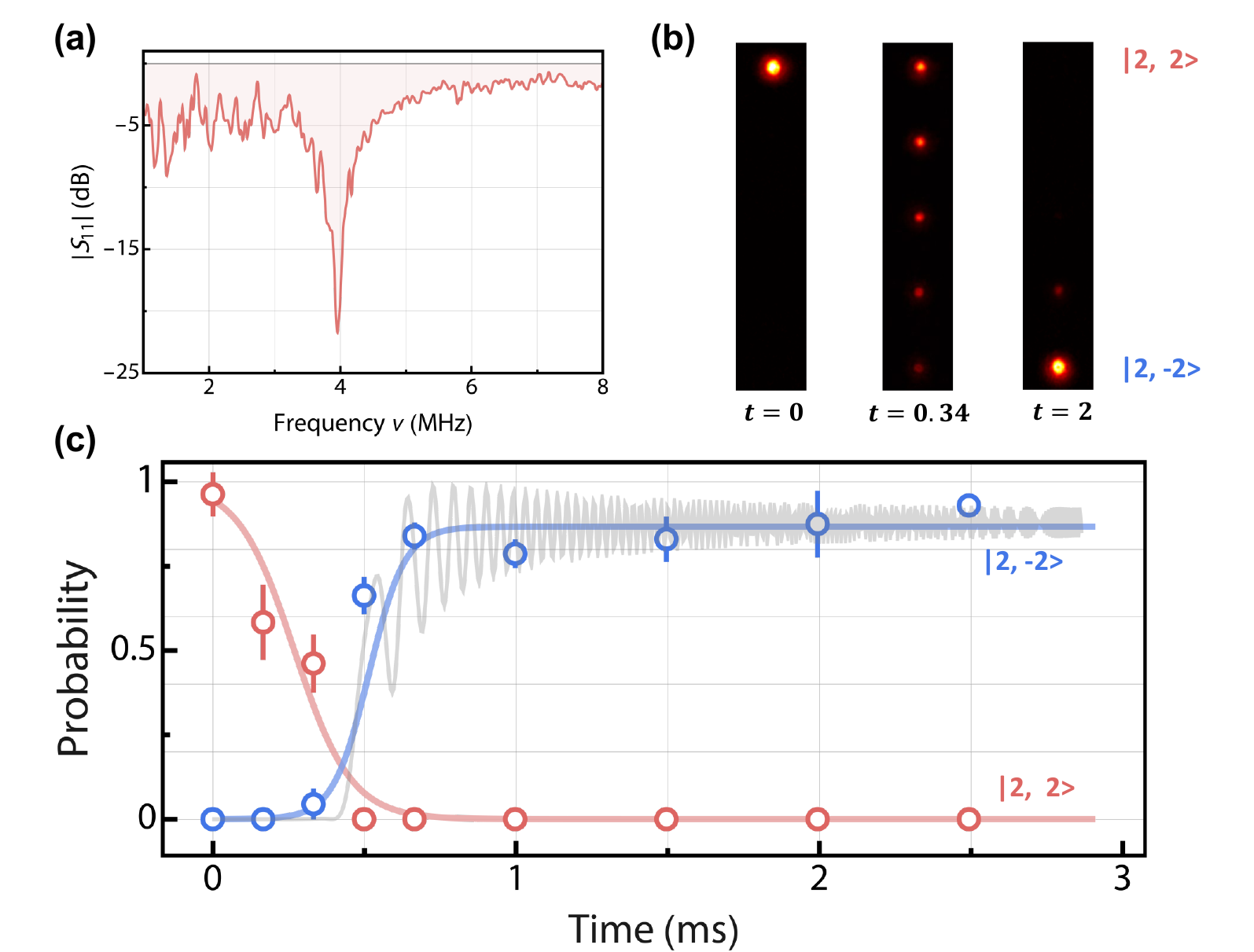}
  \caption{\label{fig5}Measurement of the reflection coefficient and Landau-Zener tunneling using a narrowband emitter based on a CTN design.
  (a) The reflection coefficient $\abs{S_{11}}$ reaches a minimum near 4\,MHz, confirming resonance. Low-pass behavior persists below this frequency.
  (b)(c) Population transfer from $\ket{2,2}$ (red circles) to $\ket{2,-2}$ (blue circles), measured using the Stern-Gerlach method. Data are normalized to the initial atom number. Error bars represent the standard deviation. The gray line shows the full numerical simulation, while solid lines represent smoothed sigmoid-like fits $\tilde{S}(t)$, used to extract the smoothed population dynamics. A vertical bias field (aligned with gravity) is linearly ramped down over \unitqty{2.9}{\milli\second}, while the radio frequency is fixed at 3.91\,MHz and the input power is set to 0.1\,dBW. The magnetic gradient is adjusted to balance the gravitational force on the $\ket{2,2}$ state. Horizontal confinement is provided by a weak crossed optical dipole trap.}
\end{figure}
At such low frequencies, our design easily achieves both impedance matching and a high peak current without requiring an optional capacitor. In this regime, matching generally yields a larger $Q_L$, hence the peak current scales approximately as $I_{\rm peak}\propto Q_L$. For example, in our narrowband design, $L_{\rm C}$ increases by less than a factor of two, whereas the resonant frequency is about one-fifth of the broadband case, resulting in at least a 2.5-fold increase in the peak current. Fig.~\ref{fig5}(a) shows the measured reflection coefficient at the input port of the network, which reaches approximately -22\,dB near \unitqty{4}{\mega\hertz}.

Near the optimal emission frequency of the coil, only a weak bias field is required to bring the hyperfine sublevels into resonance. Since the quadratic Zeeman shift is negligible, the spacing between adjacent $m_F$ states remains nearly equal. A single RF field can then couple multiple sublevels simultaneously, making it difficult to measure the Rabi frequency between two isolated states. We therefore estimate it experimentally via Landau-Zener tunneling. We first prepare a Rb Bose-Einstein condensate in the $\ket{2,2}$ state, with a condensate fraction slightly below 50\%. The RF field is then abruptly turned on and maintained at a fixed frequency with an input power of 0.1\,dBW, while the bias field is linearly ramped down within \unitqty{2.9}{\milli\second}. 
This process effectively sweeps the radio frequency detuning across the $\ket{2,2} \rightarrow \ket{2,1}$ resonance, from an initial red detuning of approximately \unitqty{15}{\kilo\hertz} to a final blue detuning of about \unitqty{105}{\kilo\hertz}. The population evolution of the initial state $\ket{2,2}$ and the final state $\ket{2,-2}$ during this process is shown in Fig.~\ref{fig5}(b)(c), with the final transfer rate exceeding 90\,\%.

By comparing the experimental results with the simulated averaged evolution of Landau-Zener tunneling (solid lines in Fig.~\ref{fig5}(c)), we estimate a Rabi frequency of about \unitqty{9}{\kilo\hertz} (see Appendix~\ref{appendixSecC}). Additionally, approximately 80\% of the atoms are transferred to the $\ket{2,-2}$ state within \unitqty{1}{\milli\second}, with no significant decrease in density. 
Moreover, the narrowband design can be flexibly optimized for any frequency below 30\,MHz, enabling high $Q_L$ and enhanced current depending on specific experimental requirements. This makes it well suited for a variety of RF-based techniques used to control Zeeman sublevels.

\section{conclusions}
In this paper, we present CTN-based matching network designs for low-frequency (below 30\,MHz) and high-power RF systems, suitable for a variety of cold atom experiments. By treating the coil as an intrinsic inductor rather than a passive termination, we developed both broadband and narrowband versions of the matching network.
The broadband configuration, validated in a Bose-Fermi mixture evaporation experiment, provided a smoothly varying current that greatly simplified evaporation timing control. The narrowband one, tested in a Landau-Zener experiment, demonstrated its potential for achieving high Rabi frequencies and rapid control, with applications in Rabi pulse-based spectroscopy\cite{Huang2016,Xu2018,Xu2019} and interferometry\cite{Sadgrove2013,Ren2023,Tiengo2025}. 

To overcome the spatial limitations posed by the metallic chamber, we designed a double-coil mount that maximizes usable space near the viewport while mitigating RF electromagnetic interference. This design offers robust and precise frequency matching, ensuring that the emission frequency remains stable, unaffected by environmental factors or wiring complexities. Beyond typical alkali-metal experiments, this design is especially well-suited for more complex experimental environments, such as cold atom experiments on the space station.

\begin{acknowledgments}
  We thank Xinyi Huang and Mingcheng Liang for insightful comments and discussions. We also thank Pengju Zhao for assistance with the cold atom experiments. This work was supported by National Key Research and Development Program of China (2021YFA1400900), the National Natural Science Foundation of China (Grants No. 12425401 and No. 12261160368), the Innovation Program for Quantum Science and Technology (Grant No. 2021ZD0302000), and the Shanghai Municipal Science and Technology Major Project (Grant No. 2019SHZDZX01).
\end{acknowledgments}

\appendix

\section{Properties of the L\text{-}type matching network (LTN)} \label{appendixSecA}
In the main text, we compared the characteristics of the conventional LTN and the CTN. We provide details on the LTN in this section. When the network in Fig.~\ref{fig1}(b) is tuned to a perfect resonance, the conjugate matching condition reads\cite{Balanis2016,Coleman2005}:
\begin{align}\label{AppendixA_eq1}
  R_{\rm S} - \frac{1}{\dsi\omega C_1} = \frac{1}{\dsi\omega C_2} \parallel (R_{\rm C} + \dsi\omega L_{\rm C}).
\end{align}
Assuming the coil resistance $R_{\rm C}$ is small compared to its reactance, we expand the solution $C_{2}$ to first order in $R_{\rm C}$, and find that the approximate resonant frequency under this loading condition is
\begin{align}
  \omega_0 \approx \frac{R_{\rm S}^{1/4}}{\sqrt{L_{\rm C}C_2 \qty(R_{\rm C}+R_{\rm S})^{1/2}}}.
\end{align}
In the absence of a source (i.e., the output circuit consisting of $C_{1,2}$ and coil only), the unloaded $Q_U$ of the RLC resonator is given by:
\begin{align} \label{AppendixA_eq3}
  Q_U&=\frac{1}{R_{\rm C}}\sqrt{\frac{L_{\rm C}}{C_1+C_2}}.
\end{align}
This represents the intrinsic quality factor of the passive circuit, excluding the effect of the source resistance. Using the approximate values of $C_{1,2}$ obtained from Eq.~\ref{AppendixA_eq1} in the small $R_{\rm C}$ limit, we further obtain the expression:
\begin{align} \label{AppendixA_eq4}
  Q_U\approx \frac{\omega_0 L_{\rm C}}{R_{\rm C}}-\frac{\sqrt{R_{\rm S}R_{\rm C}}}{2\omega_0 L_{\rm C}}.
\end{align}
When the circuit is connected to a source with internal resistance $R_{\rm S}$, the source introduces additional power loss, which can be described by the source-limited quality factor $Q_S$. Under the matched condition, the total loaded quality factor of the LTN is then given by\cite{Pozar2012}:
\begin{align} \label{AppendixA_eq5}
  Q_L=Q_S\parallel Q_U=\frac{Q_U}{2}.
\end{align}
Comparing Eqs.~\ref{AppendixA_eq4} and \ref{AppendixA_eq5} to Eq.~(\ref{eq3}) in the main text, we find that the LTN behaves as a loaded series RLC circuit. In atomic experiments, the coil reactance is typically much larger than its resistance, resulting in a large $Q_L$ and thus a narrow matching bandwidth in the low-frequency regime. By contrast, the CTN design retains a similar loaded $Q_L$ characteristic of a parallel RLC circuit, but with an additional impedance transformation. This allows the CTN to maintain current enhancement while providing a broader bandwidth.

Furthermore, in the ultra-low-frequency limit, the large impedance of the capacitors in the LTN significantly suppresses current flow. This confirms that the conventional LTN lacks the low-pass behavior that is desirable for broadband evaporative cooling.

\section{Range of validity for Eq.~(\ref{eq1})} \label{appendixSecB}
The simple resonant condition requires two conditions to be met. First, the virtual load resistance $R_{\rm L}$ must be much greater than $1/\qty(\omega C_{1,2})$. Second, $R_{\rm C}$ must be much smaller than $\omega L_{\rm C}$. The former condition limits the frequency to a relatively low range, with approximately 30\,MHz being a safe upper limit empirically in cold atom experiments. The latter condition significantly affects the resonant frequency estimation at ultra-low frequencies (around a few MHz). In fact, since $R_{\rm in}$ is determined by the parallel combination of $\tilde{R}_{\rm C}$ and $\tilde{R}_{\rm L}$, if the frequency-dependent $\tilde{R}_{\rm C}=\qty[R_{\rm C}^2 + \qty(\omega L_{\rm C})^2] / R_{\rm C}$  satisfies $\tilde{R}_{\rm C} < R_{\rm S} $, it will inevitably result in $R_{\rm in} < R_{\rm S}$ , thereby violating real-part impedance-matching condition in Eq.~\ref{eq1}. For the parameters used in Fig.~\ref{fig1}(c)(d), when the frequency drops below approximately 11.1\,MHz, the condition $\tilde{R}_{\rm C} < R_{\rm S}$ arises, which prevents exact resonance from being achieved. However, this lower-frequency limit can always be reduced by increasing the inductance of the coil. 
In contrast, adding an optional capacitor $C_{\rm op}$ modifies the reactance in the coil branch from $\omega L_{\rm C}$ to $\omega\qty[L_{\rm C}-1/\qty(\omega^2C_{\rm op})]$, effectively shifting the red dashed line to a new condition $L_{\rm crit^\prime}\approx L_{\rm crit}+1/\qty(\omega^2C_{\rm op})$, where $L_{\rm crit^\prime}(L_{\rm crit})$ is the critical inductance required for exact impedance matching after (before) adding $C_{\rm op}$. The effect of this shift is shown in Fig.~\ref{fig2}(a)(b).

\section{Estimation of the Rabi Frequency from Landau-Zener Tunneling} \label{appendixSecC}
\begin{figure}
  \includegraphics[scale=0.28]{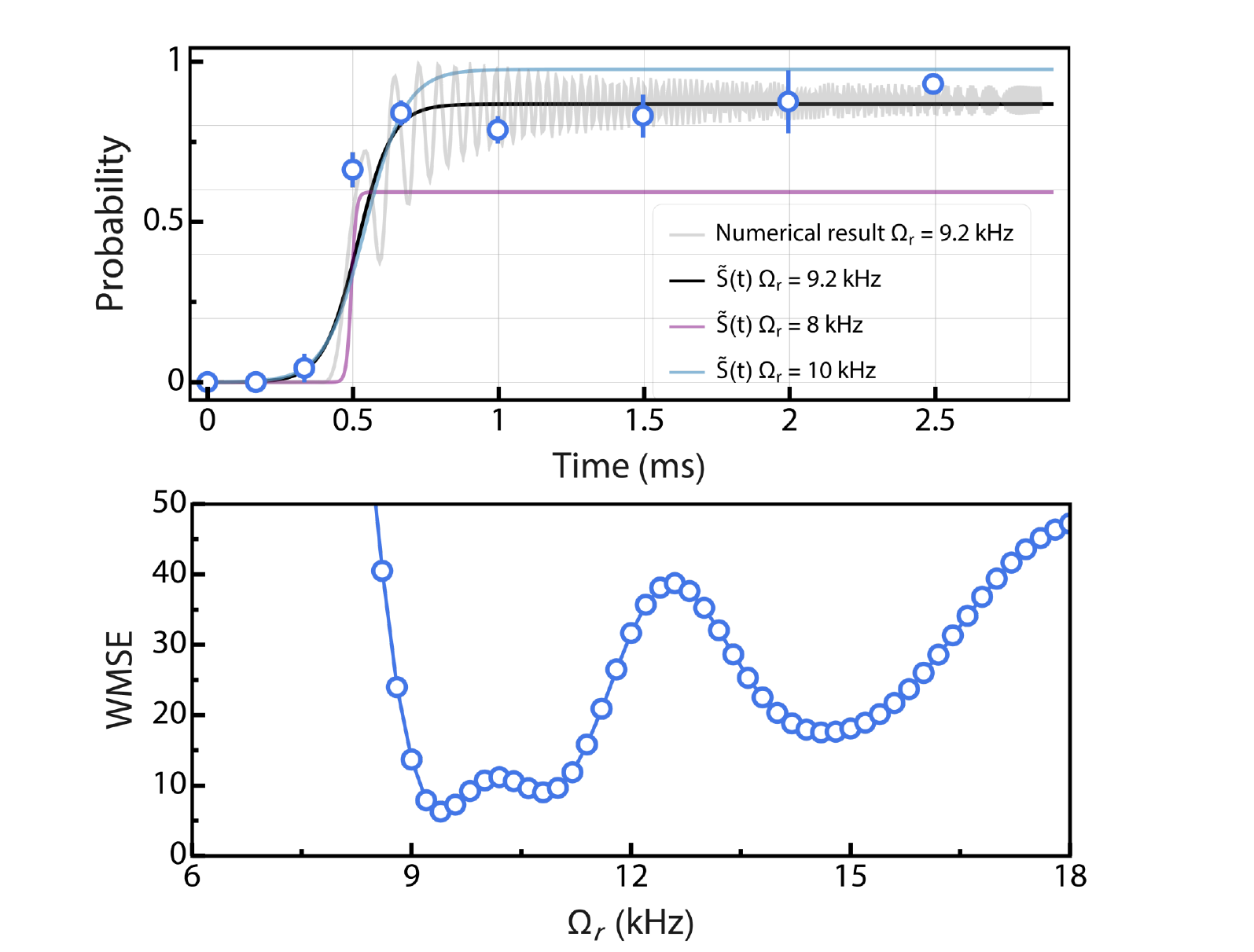}
  \caption{\label{FigSM3} 
  Estimation of the effective Rabi frequency using a five-level model. The top panel shows the simulated population transfer to $\ket{2,-2}$ under different values of $\Omega_r$. Fast oscillations are smoothed using a sigmoid-like function $\tilde{S}(t)$ to extract the averaged evolution. The bottom panel shows the weighted mean squared error (WMSE) between $\tilde{S}(t)$ and the experimental data, with the minimum yielding the optimal estimate of $\Omega_r$.}
\end{figure}
The Landau-Zener tunneling between Zeeman sublevels of the $\ket{F = 2}$ manifold can be simulated using a five-level model in the rotating frame. We choose the basis $\ket{\psi} = \sum_{i=-2}^{2} \ket{2, m_F = i}$. The lab-frame Hamiltonian is given by
\begin{align}
  H(t) = \sum_{i=-2}^{2} \varepsilon_i(t) \ketbra{i}{i} + \sum_{j=-2}^{1} \left( \frac{\Omega_r}{2} e^{\dsi \omega_{\rm rf} t} \ketbra{j}{j+1} + \mathrm{h.c.} \right),
\end{align}
where $\varepsilon_i(t) = \varepsilon_{i,0}(t) + \varepsilon_{i,1}$ denotes the $m_F$-dependent energy shifts due to external bias fields. In this simulation, we include only the leading-order linear Zeeman effect in $\varepsilon_{i,0}(t)$, which varies with time through the sweep of magnetic field. The quadratic Zeeman shift $\varepsilon_{i,1}$ is treated as constant, since the total field change during the sweep is small (approximately $\delta B \approx 170\,{\rm mG}$).
Here, $\Omega_r$ denotes the standard Rabi frequency\cite{steck2023} to be estimated. In the rotating frame, the Hamiltonian becomes
$H_{\rm rot}(t) = U H(t) U^\dagger + \dsi \partial_t U U^\dagger,$
with $U = \exp\left( \dsi \omega_{\rm rf} t \sum_i i \ketbra{i}{i} \right)$. The time evolution is simulated numerically as $ \ket{\psi(t)} = U(t,0)\ket{\psi(0)} $, with the initial state $\ket{\psi(0)} = \ket{2, 2}$. The time-evolution operator is given by $U(t,0) \equiv \mathcal{T} \exp\left(-\dsi \int_0^t H_{\rm rot}(t')\,dt'\right)$, where $\mathcal{T}$ denotes the time-ordering operator.
The numerical results for final state $\ket{2, -2}$ are shown in Fig.~\ref{FigSM3}. The averaged time evolution is fitted using a smoothed sigmoid-like function of the form:
\begin{align}
\tilde{S}(t) = \frac{1}{a + b e^{-\lambda (t - t_0)}},
\end{align}
and the fit quality is evaluated using the weighted mean squared error (WMSE), defined as $\chi = \frac{1}{N} \sum_i (p_i - \tilde{S}(t_i))^2/\sigma_i^2$,
where $N$ is the number of data points, $p_i$ is the measured probability in $\ket{2, -2}$, and $\sigma_i$ is the corresponding uncertainty.

Due to the absence of an analytical solution for the complete Landau\text{-}Zener dynamics under our scheme, this residual-based method offers a practical way to estimate the Rabi frequency. More importantly, we observe that further reduction of the RF power in the experiment prevents the completion of the transfer process, indicating that this setting corresponds to the minimum Rabi frequency required to achieve nearly complete transfer from $\ket{2, 2}$ to $\ket{2, -2}$ within the given ramp sequence. 
Thus, the estimated value of $\Omega_r$ corresponds to the first dip in the WMSE plot, with a range of approximately $9.0\text{--}9.5$\,kHz.

\end{document}